\documentclass[conference]{IEEEtran}
\IEEEoverridecommandlockouts
\usepackage{cite}
\usepackage{amsmath,amssymb,amsfonts}
\usepackage{textcomp}
\usepackage{xcolor}
\def\BibTeX{{\rm B\kern-.05em{\sc i\kern-.025em b}\kern-.08em
		T\kern-.1667em\lower.7ex\hbox{E}\kern-.125emX}}
\usepackage[normalem]{ulem}
\useunder{\uline}{\ul}{}
\usepackage{longtable}
\usepackage{amsmath}
\usepackage{graphicx,epstopdf}
\usepackage{cite}
\usepackage{caption}
\usepackage{amsfonts}
\usepackage{array}
\usepackage{bm}
\usepackage{framed}
\usepackage{tikz}
\usepackage{algorithmic}
\usepackage[]{algorithm}
\usepackage[margin=2cm]{geometry}
\usepackage{tabularx}
\usepackage{amsthm}
\usepackage{amssymb}
\usepackage{subcaption}
\usepackage{color} 
\usepackage[numbers,sort&compress]{natbib}
\usepackage{multirow}
\usepackage{alltt}
\usepackage[normalem]{ulem}
\usepackage[utf8]{inputenc}

\usepackage{bigstrut}
\usepackage{booktabs}
\usepackage{alltt}
\usepackage{multirow,booktabs}
\usepackage{multirow} 

\theoremstyle{remark}

\newtheorem{theorem}{Theorem}
\newtheorem{lemma}{Lemma}

\definecolor{mycolor1}{rgb}{0.97255,0.97255,0.97255}%
\newcommand{\distas}[1]{\mathbin{\overset{#1}{\kern\z@\sim}}}%

\def \Q {{\mathbf Q}}
\def \EE {{\mathbb E}}
\usepackage{mathtools}

\def \s {{\mathbf s}}
\def \q {{\mathbf q}}
\def \g {{\mathbf g}}

\def \cO {{\mathcal O}}
\def \cX {{\mathcal X}}

\def \Rn {{\mathbb R}}

\def \xh {{\hat{\x}}}

\def \dF {{\nabla}}

\def \x {{\mathbf x}}
\def \xs {{\x^{\star}}}
\def \w {{\mathbf w}}
\def \y {{\mathbf y}}
\def \h {{\mathbf h}}
\def \z {{\mathbf z}}
\def \p {{\mathbf p}}

\def \pr {{\text{prox}}}
\newcommand{\norm}[1]{\ensuremath{\left\|#1\right\|}}						

\usepackage{mathtools}
\newcommand{\genFkl}[1]{\expandafter\newcommand\csname k#1\endcsname{{\mathfrak #1}}}

\renewcommand{\vec}[1]{{\mathbf{#1}}}
\newcommand{\genLatinVec}[1]{\expandafter\newcommand\csname v#1\endcsname{{\vec #1}}}

\newcommand{\mc}[1]{{\mathcal{#1}}}
\newcommand{\mb}[1]{{\mathbb{#1}}}
\newcommand{\genLatinVecU}[1]{\expandafter\newcommand\csname v#1\endcsname{{\vec #1}}}
\def\mydefgreek#1{\expandafter\def\csname v#1\endcsname{\text{\boldmath$\mathbf{\csname #1\endcsname}$}}}
\def\mydefallgreek#1{\ifx\mydefallgreek#1\else\mydefgreek{#1}%
	\lowercase{\mydefgreek{#1}}\expandafter\mydefallgreek\fi}
\mydefallgreek {alpha}{beta}{gamma}{delta}{epsilon}{varepsilon}{zeta}{eta}{theta}{vartheta}{iota}{kappa}{lambda}{mu}{nu}{xi}{omicron}{pi}{rho}{varrho}{sigma}{tau}{upsilon}{phi}{varphi}{chi}{psi}{omega}\mydefallgreek

\def\mydefugreek#1{\expandafter\def\csname v#1\endcsname{\text{\boldmath$\mathbf{\csname #1\endcsname}$}}}
\def\mydefallugreek#1{\ifx\mydefallugreek#1\else\mydefugreek{#1}%
	\lowercase{\mydefugreek{#1}}\expandafter\mydefallugreek\fi}
\mydefallugreek{Gamma}{Delta}{Theta}{Lambda}{Xi}{Sigma}{Upsilon}{Phi}{Psi}{Omega}\mydefallugreek

\newcommand{\vzero}{\vec{0}}

\newcommand{\bc}[1]{\left\{{#1}\right\}}
\newcommand{\br}[1]{\left({#1}\right)}
\newcommand{\bs}[1]{\left[{#1}\right]}

\newcommand{\ip}[2]{\left\langle{#1},{#2}\right\rangle}

\newcommand{\E}[1]{\mb{E}\bs{{#1}}}
\newcommand{\Ee}[2]{\mb{E}_{{#1}}\bs{{#2}}}

\begin{document}
	
	%
	%
	\title{Optimizing QoS for Erasure-Coded Wireless Data Centers}
	
	\author{\IEEEauthorblockN{Srujan Teja Thomdapu, Ketan Rajawat}
		\IEEEauthorblockA{Department of Electrical Engineering, Indian Institute of Technology Kanpur, India 208016\\
			Email: \{srujant, ketan\}@iitk.ac.in}
	}

	\maketitle
	
	
	\begin{abstract}
	Cloud computing facilitates the access of applications and data from any location by a distributed storage system. Erasure codes offer better a data replication technique with reduced storage costs for more reliability. This paper considers the erasure-coded data center with multiple servers in a wireless network where each is equipped with a base-station. The cause of latency in the file retrieval process is mainly due to queuing delays at each server. This work puts forth a stochastic optimization framework for obtaining the optimal scheduling policy that maximizes users' quality of service (QoS) while adhering to the latency requirements. We further show that the problem has non-linear functions of expectations in objective and constraints and is impossible to solve with traditional SGD like algorithms. We propose a new algorithm that addresses compositional structure in the problem. Further, we show that the proposed algorithm achieves a faster convergence rate than the best-known results. Finally, we test the efficacy of the proposed method in a simulated environment.
	\end{abstract}
	
	\section{Introduction}
	Increased cloud computing services' requirements are making companies look at distributed storage systems where efficient data replication techniques are implemented for more reliability. Data redundancy helps in providing more alternatives to the clients in case of node failures. Most of the cloud-based companies such as Facebook \cite{asteris2013xoring}, Microsoft \cite{huang2012erasure}, and Google \cite{fikes2010storage} have found the erasure coding technique as the most prominent solution to reduce storage cost compared to other techniques \cite{weatherspoon2002erasure,dimakis2010network}. 
	
	Besides being a promising idea, erasure-coded storage systems face a troublesome issue, which is the delay that occurs while downloading files from the data center by users. There have been much fewer works that have studied quantitative analysis on queuing delays. Most of the existing works \cite{aguilera2005using,li2006adaptive} instead have focused on designing efficient data systems without making much effort into analyzing the queues that appear at each data server. Hence, many researchers have turned their attention towards latency analysis these days making it as an active area \cite{schurman2009user}. There have been many recent attempts in obtaining latency bounds in erasure-coded storage systems by proposing various scheduling policies include `block-one-scheduling' \cite{huang2012codes}, `fork-join queue' \cite{joshi2014delay}, `probabilistic scheduling' \cite{xiang2016joint}. It has been shown in \cite{xiang2016joint} that, probabilistic scheduling policy provides an upper bound on average latency of erasure-coded storage systems for arbitrary erasure codes with M/G/1 queues at data servers. The policy entertains scheduling file requests to all the possible servers. Analysis of erasure-coded storage systems has been extended to video streaming case \cite{al2018video} where an optimized service has been proposed that maximizes the quality of experience (QoE) for users. More precise latency analysis is pursued in \cite{al2018video} by assuming service time distribution as exponential.   
	
	 \begin{figure*}
		\setcounter{subfigure}{0}
		\begin{subfigure}{\columnwidth}
			\includegraphics[width=0.9\linewidth, height = 0.65\linewidth]
			{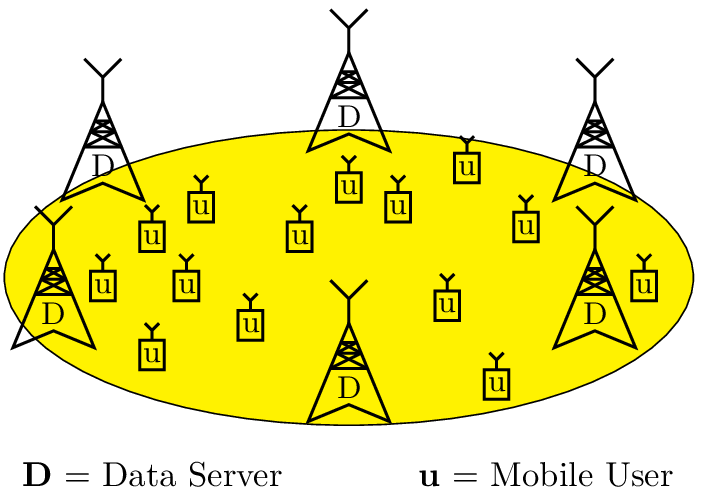}
			\caption{Data Center} \label{sysFig}
		\end{subfigure}
		\begin{subfigure}{\columnwidth}
			\includegraphics[width=0.9\linewidth,height = 0.55\linewidth]
			{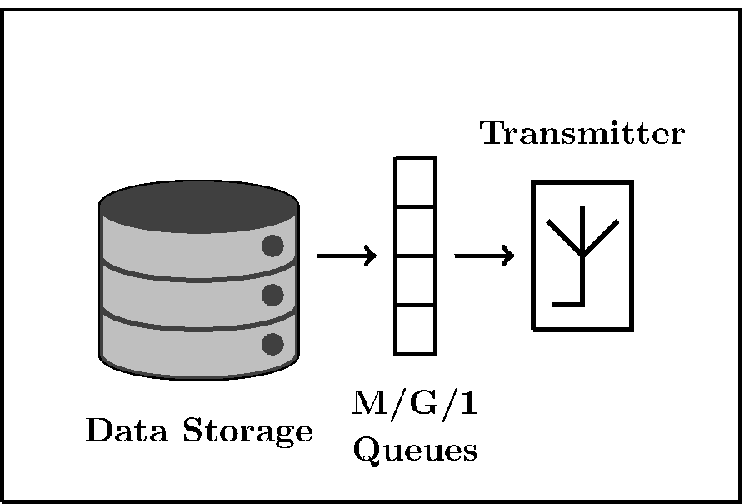}
			\caption{Entities in Data Server}
			\label{dataServer}
		\end{subfigure}
		\caption{System Model}
		\label{sysModFig}
	\end{figure*}	   
	
	In this paper, we consider distributed erasure-coded storage systems in wireless networks where each server is equipped with a multi-antenna base-station that is capable of wireless transmissions. Specifically, we formulate the stochastic optimization problem to find an optimal scheduling policy that maximizes users' quality of service (QoS) while adhering to queuing delay and other deterministic constraints. Since the file transfer medium is considered wireless, we can no longer assume exponential service time distributions due to the random fading channels between user and data center. However, the classical approach is not suited due to the difficulty in evaluating closed form expressions for first and second order moments of service times with the presence of exogenous variables \cite{lee2009average}. Recently, the authors in \cite{thomdapu2019optimal} have considered the design of queuing systems from a stochastic optimization perspective where the queues have general service time distributions. By applying ideas of \cite{thomdapu2019optimal}, we show that the formulated problem has at least one of the objective and constraints has non-linear function of expectations. Hence, is not solvable with the existing SGD like first-order methods as they require unbiased estimates of (sub) gradients (see \cite{benveniste2012adaptive}).  
	
	Much recent work in \cite{wang2017stochastic} has presented a first-order method that deals with the non-linear functions of expectations via the stochastic compositional gradient descent (SCGD) algorithm. Since finding true (sub) gradients is not possible due to the composition structure, the SCGD algorithm adopts a quasi-gradient approach by estimating the approximated (sub) gradients. Accelerated version of SCGD is later proposed in \cite{wang2017accelerating}. The structure of the problem is, however, considered to be unconstrained in \cite{wang2017stochastic,wang2017accelerating}. Recently, the CSCGD algorithm has been proposed in \cite{thomdapu2019optimal} to solve constrained stochastic compositional problems. Exploring the ideas from \cite{wang2017accelerating,thomdapu2019optimal}, we propose a new algorithm that solves constrained stochastic compositional problems with a faster convergence rate.
	
	The rest of the paper is organized as follows. Sec. \ref{sys-mod} details the system model and problem formulations. The algorithm and theoretical guarantees for it are provided in Sec. \ref{sec-scgd}. Later, we evaluate the performance of the proposed method in Sec. \ref{sim}. Finally, we conclude our paper in Sec. \ref{concl}. 	
	
	\section{System Model And Problem Formulation} \label{sys-mod} 
	We consider a data center, as shown in Fig.\ref{sysModFig}, with $M$ servers denoted by $\mc{M}$ that are placed in an area that is populated by users with mobile devices. Each server contains a base-station of multiple antennas that serve users through wireless transmissions, and is also capable of storing data (see Fig. \ref{dataServer}). We do not consider any interference management schemes, and hence we assume all the concurrent data transmissions by servers are transmitted in orthogonal channels. There are $N$ trending files of current interest, which are picked out of millions and may be delivered to the users. These selected files are the most popular ones, and they earned the placement opportunity at servers. The data objects are represented as $Z_i$ $\forall i \in \bc{1,...,N}$. Users can request any of these popular $N$ files from the data center. 
	
	\subsection{Coding}
	Each file $Z_i$ is divided into $k_i$ number of fixed-size chunks and then encoded using $(n_i,k_i)$ Maximum Distance Separable (MDS) erasure code to generate $n_i$ distinct chunks for a file. These coded chunks are denoted as $C_i^{(1)}, ..., C_i^{(n_i)}$. The encoded chunks are stored in distinct $n_i$ servers for delivery. The servers that store the chunks corresponding to the file $i$ are denoted by a set $\mc{S}_i$ such that $\mc{S}_i \subset \mc{M}$, and $|\mc{S}_i| =n_i$. A typical MDS erasure code $(n_i,k_i)$ is in such a way that $n_i>k_i$, with the redundancy factor $n_i/k_i$. Hence any subset of $k_i$-out-of-$n_i$ coded chunks can reconstruct the original file. For example, a simple replication of a file at $n$ servers is nothing but using $(n,1)$ erasure code. We assume a centralized system that knows file placement information and schedules all the user requests to different servers. Hence, when a file is requested, the request goes to a set $\mc{A}_i$ storage nodes where $\mc{A}_i\subset\mc{S}_i$, and $|\mc{A}_i| = k_i$. Each server maintains a FIFO queue to serve the users, as shown in Fig.\ref{dataServer}. When a file $i$ is requested, all the chunks for other file requests have not yet been served are waiting. 
	
	\subsection{Policy}
	The central system distributes the users' requests based on the file availabilities. It has to pick the optimal choice out of many options to reduce the latency to schedule a request of file $i$ to the $k_i$ server. We use the Probabilistic Scheduling policy as proposed in \cite{xiang2016joint} That allows the choice of every possible subset of $k_i$ nodes with a certain probability. Once a request for a file $i$ arrives, the central system randomly distributes the $k_i$ chunk requests to a set of nodes $\mc{A}_i$ with predetermined probabilities $P(\mc{A}_i)$. Then each server maintains a local queue containing all the chunk requests to finally be transmitted by the base-station. According to the Probabilistic scheduling policy, the feasible probabilities $P(\mc{A}_i)$ exist when the following conditions are satisfied.
	\begin{align} 
	\sum_{j=1}^{m} \pi_{ij} = k_i \forall i \hspace{2mm} \pi_{ij} = 0  \text{ If } j \notin \mc{S}_i, \label{polfea}
	\end{align}
	where $\pi_{ij}$ is the conditional probability of selecting the node $j$ for the request $i$.
	
	When we consider a simple case where only a file is placed in the system, ideally, in that case, the servers with which more users have good channels need to be scheduled. In other words, the probabilities of choosing such servers are high. The current setting however, is much more complicated. Although the files are more popular in the current scenario, the demographic preferences are unknown at this stage. For example, users that reside at various geographical locations may have different priorities towards these files. Hence to reduce the latency, we try to incorporate all of these scenarios while formulating the problem.   
	
	\subsection{Queuing Model}
	We assume the file requests follow the Poisson process with a known rate of $\lambda_i$. We can think of these rates $\lambda_i$ as file popularities. Hence the arrival of chunk requests at node $j$ follow the Poisson with rate $\Lambda_j = \sum_{i}\lambda_i\pi_{ij}$. The chunk service time distribution is unknown and is subject to the fading channel's random behavior, thus making the queuing models M/G/1. Let $\xi_{ij}$ represents the channel constant corresponding to the user who requested file $i$, and $p_j$ is the power allocated to the transmissions from the node $j$. If the length of coded packet $C_i^{\mc{S}_i}$ is $L$, then the random service time $X_j$ for the packets that leave from node $j$ can be written as
	\begin{align*}
	X_j = \frac{L}{b_j(p_j,\xi_{ij})} \hspace{3mm} \text{with probability} \hspace{3mm} \frac{\pi_{ij} \lambda_i}{\Lambda_j} \hspace{2mm} \forall i,
	\end{align*}  
	where $b_j(p_j,\xi_{ij}) = B_j\log(1+p_j\xi_{ij})$ is Shannon's capacity, and $B_j$ is the bandwidth of the channel. Now the first-order moment for service time can be found as
	\begin{align*}
		\E{X_j} &= \E{\E{X_j\lvert \xi}} \\
		& = L\Ee{\xi}{\sum_{i}^N\frac{\pi_{ij} \lambda_i}{\Lambda_j b_j(p_j,\xi_{ij})}} \\
		& = L\sum_{i}^N\frac{\pi_{ij} \lambda_i}{\Lambda_j}\Ee{\xi}{\frac{1}{b_j(p_j,\xi_{ij})}}
	\end{align*} 
	where $\Ee{\xi}{.}$ is expectation w.r.t to the only random variable $\vxi$. Similarly, we can also find second-order moment as 
	\begin{align*}
		\E{X_j^2} = L^2\sum_{i}^N\frac{\pi_{ij} \lambda_i}{\Lambda_j}\Ee{\xi}{\frac{1}{b^2_j(p_j,\xi_{ij})}}
	\end{align*}
	Now the average queuing delay can be calculated using the Pollaczek-Khinchin (P-K) formula as 
	\begin{align}
	 W_j (p_j,\pi_{ij})&=  \frac{\Lambda_j\E{X_j^2}}{2(1-\Lambda_j\E{X_j})} \nonumber\\
	&= \frac{L^2\sum_{i}^N\pi_{ij} \lambda_i\Ee{\xi}{\frac{1}{b^2_j(p_j,\xi_{ij})}}}{2\br{1-L\sum_{i}^N\pi_{ij} \lambda_i\Ee{\xi}{\frac{1}{b_j(p_j,\xi_{ij})}}}}. \label{qD}
	\end{align}
	The complicated looking expression in \eqref{qD} restricts from finding closed form for any known distributions of channel constants $\{\xi_{ij}\}$.
	
	\subsection{Problem Formulation}
	By assuming the information of file placement in servers is known, we formulate the problem to obtain an optimal policy that enhances the QoS of users with the following utility function 
	\begin{align}
	U(\p) = \sum_{j=1}^M \psi\br{\E{\sum_{i}b_j(p_j,\xi_{ij})}}, \label{qos}
	\end{align} 
	where $\psi(.)$ is any function that include such as linear, log utility functions. A simple observation tells us that the function in \eqref{qos} is convex. By imposing constraints on queuing delay, the problem finally can be written as
	\begin{subequations}\label{appProb}
	\begin{align} 
	\max_{\p,\Pi} & ~U(\p) \label{proB}\\
	\text{s.t }& ~W_j(p_j,\pi_{ij}) \leq D_j \text{ }\forall j \label{consc}\\
	& \sum_{j=1}^M p_j \leq P, \text{ } \eqref{polfea}. \label{detcon}
	\end{align}
	\end{subequations}
	The objective in \eqref{proB}, and constraint in \eqref{consc} functions are stochastic in nature while the constraints in \eqref{detcon} are deterministic which are simpler to project onto. The collection of all the random variables $\bc{\xi_{ij}}$ is represented as $\vxi$. The goal is to solve the above problem in online fashion using independent realizations $\vxi_1$, $\vxi_2$, $...$ that are revealed sequentially. The constraints in \eqref{consc} are expressed as non-linear functions of sample probabilities. Hence, the existing first-order methods do not apply here due to unbiased gradient estimates' requirement to objective and constraint functions. The constraints in \eqref{consc} are convex functions (see \cite[Appendix C]{thomdapu2019optimal}) thus making the whole problem in \eqref{appProb} convex. To the best of our knowledge, optimizing QoS in erasure-coded wireless data centers has not been considered in the existing literature. Algorithmic details to solve \eqref{appProb}, and the theoretical convergence guarantees for it are provided in the subsequent section. 

\section{Accelerated Stochastic Compositional Gradient Descent for Constrained Problems}\label{sec-scgd}
Consider the more general constrained stochastic optimization problem 
\begin{align} \label{mainProb}
\x^\star = \arg&\min_{\x \in \cX} ~ f(\E{\g(\x,\vxi)}) + R(\x)  \nonumber \\
&\text{s.t. }  \hspace{2mm} \q(\E {\h(\x, \vxi)})\leq \vzero \tag{$\mathcal{P}$}
\end{align}  
where the expectation is taken with respect to $\vxi$. Here, $f:\Rn^m\rightarrow \Rn$, $\g:\Rn^n\times \Rn^k\rightarrow\Rn^m$, $\h:\Rn^n\times \Rn^k\rightarrow\Rn^d$, and $\q:\Rn^d \rightarrow \Rn^J$ are continuous functions. The penalty function $R(\x) : \Rn^n \rightarrow \Rn \cup \bc{+\infty}$ is an extended real-valued closed convex function which is allowed to be non-smooth. The problem can have simple deterministic constraints as in \eqref{detcon} which can be added in $R(\x)$. It can be easily verified that the problem formulated in \eqref{appProb} is a spacial case of \eqref{mainProb}. Since the distribution of $\vxi$ is unknown, the expectations appearing in \eqref{mainProb} cannot be evaluated in closed-form. Motivated by classical stochastic approximation methods, the goal is to solve \eqref{mainProb} in an online fashion using only independent realizations $\vxi_1$, $\vxi_2$, $\ldots$ that are revealed sequentially. This section details the proposed algorithm for solving \eqref{mainProb} and provides the corresponding convergence rates. For the sake of brevity, we define $F(\x):=f(\E{\g(\x,\vxi)})$, $\Q(\x):=\q(\E{\h(\x,\vxi)})$.

\subsection{Assumptions}\label{asm}
We begin with discussing the necessary assumptions on the functions $f$, $\g$, $q$ and $\h$. All functions $f$, $\q$, $\g$, $\h$ are continuously differentiable. Consequently, the gradients of the objective and constraint functions are well-defined, with The problem \eqref{mainProb} is a convex optimization problem and the set $\cX$ is  closed and compact, i.e., $	\sup_{\x,\x' \in \cX} \norm{\x - \x'}^2 \leq D_x < \infty$. The random variables $\vxi_1, \vxi_2,...$ are  independent and identically distributed. The functions $\g$, $\h$ are Lipschitz continuous in expectation and have bounded second order moments. The functions $f$, $\q$ are smooth and have bounded gradients. The functions $F$, $\Q$, and the inner functions $\g$, $\h$ are smooth.

\subsection{Proposed Algorithm}

\begin{algorithm}
	\caption {Accelerated Constrained Stochastic Compositional Proximal Gradient (ACSCPG)}
	\begin{algorithmic}[1]
		\STATE\textbf{Input:} $\x_1\in\mathbb{R}^n$ step sizes $\alpha_t,\beta_t,\delta_t\subset(0,1]$.
		\STATE\textbf{Initialize} $\y_1 = \z_1 = 0$, $\w_1 = \x_1$.
		\STATE\textbf{for} $t =1,2,...$
		\STATE\hspace{3mm}Observe the random variable $\vxi_t$, and update $\x_{t+1}$
		\begin{align}
		= \pr_{\alpha_tR(.)}\bigg\{\x_t - \alpha_t\dF\g(\x_t,\vxi_t)\dF f(\y_{t}) \nonumber\\
		- \delta_t\dF \h(\x_t,\vxi_t)\dF \q(\z_{t})\dF\ell(\q(\z_{t}))\bigg\} \label{xup}
		\end{align}
		\STATE\hspace{3mm}Observe the random variable $\vxi_{t+1}$, and  update the axillary iterates as
		\begin{align}
		\w_{t+1} &= \br{1-\frac{1}{\beta_t}}\x_t + \frac{1}{\beta_t}\x_{t+1}, \label{wup}\\
		\y_{t+1} &= \br{1-\beta_t}\y_t  + \beta_t\g\br{\w_{t+1},\vxi_{t+1}}, \nonumber \\
		\z_{t+1} &= \br{1-\beta_t}\z_t  + \beta_t\h\br{\w_{t+1},\vxi_{t+1}} \label{auxup}
		\end{align}  
		\STATE\textbf{end}
		\STATE\textbf{Output:} $\xh = \frac{2}{T}\sum_{t = T/2}^T\x_{t+1}$.	
	\end{algorithmic}
	\label{cscgda}
\end{algorithm}  

Similar to the analysis in \cite{thomdapu2019optimal}, we define a smooth and convex function $\ell(\w) = \sum_{j=1}^J\ell_j(w_j)$ where $\ell_j$ is defined as
\begin{align} \label{hub}
\ell_j(x) := \begin{cases} \tfrac{1}{2} x^2 & 0 \leq x \leq C_\ell \\
C_\ell x - \frac{C_\ell^2}{2} & x > C_\ell \\
0 & x < 0
\end{cases}
\end{align}
Gradient of the function is bounded as $\dF \ell_j(x)=\max\{x,C_\ell\}$ for $x>0$ and zero otherwise. The parameter $C_l$ in our case is the upper limit of $\max_jq_j(\E {\h(\x, \vxi)})$ for any $\x \in \cX$. The penalty function that is defined in \eqref{hub} helps in taking the iterate towards the descent direction of optimal function $F$ as well as towards the feasible region $\{\x: \Q(\x)\leq 0\}$. 

As proposed in \cite{thomdapu2019optimal}, CSCGD algorithm carries updates towards the negative direction of the approximated gradients of both objective $F(\x)$, and penalty $\ell(\Q(\x))$ functions. However, In the present case, iterates are of similar except, the steps that track $\E{\g(\xs,\xi)}$, and $\E{\h(\xs,\xi)}$. We introduce a new step in \eqref{wup} that tracks running average of the optimal solution and is used in auxiliary variable updates as shown in \eqref{auxup} and that is called extrapolation-smoothing scheme which is the main reason behind the acceleration of convergence. The complete procedure is summarized Algorithm \ref{cscgda}. Compared to the CSCGD, ACSCPG estimates the unknown quantities $\E{\g(\x,\vxi)}$, $\E {\h(\x, \vxi)}$ with faster rate. The updates in \eqref{wup}, \eqref{auxup} are carried out in way that the $\y_t$, $\z_t$ are approximately unbiased estimates of $\E{\g(\x,\vxi)}$, $\E {\h(\x, \vxi)}$. To explicitly see that, let us define the weights as
\begin{align}
\zeta_k^{(t)} = \begin{cases}
\beta _k\prod_{i = k+1}^{t}\br{1-\beta_i}& \text{ if } t>k\geq 0 \\ 
\beta_t& \text{ if } t=k\geq 0, 
\end{cases}
\end{align}
then we have the following relations
\begin{align*}
\x_{t+1} &= \sum_{k=0}^{t}\zeta_k^{(t)} \w_{k+1}, \\
\y_{t+1} &= \sum_{k=0}^{t}\zeta_k^{(t)} \g(\w_{k+1},\vxi_{k+1}), \\
 \z_{t+1} &= \sum_{k=0}^{t}\zeta_k^{(t)} \h(\w_{k+1},\vxi_{k+1})
\end{align*}
In other words, $\x_{t+1}$ is weighted average of $\{\w_{t}\}_1^{t+1}$, and $\y_{t+1}$, $\z_{t+1}$ are weighted averages of $\{\g(\w_{k+1},\vxi_{k+1})\}_1^{t+1}$, $\{\h(\w_{k+1},\vxi_{k+1})\}_1^{t+1}$. Hence as $t$ progresses, the estimates $\y_{t}$, $\z_t$ reach much nearer to the unbiased gradients of inner functions. 

\begin{table*}
	\centering
	\caption{Summary of bounds on optimality gap and constraint violation} 
	\begin{tabular}{|l|l|l|}
		\hline
		Choice of constants $a,b,c,d,e$ & Optimality gap & Constraint violation   \\ \hline
		$a = 0.9048$, $b = 0.5714$, $c = 0.7143$, $d = -0.0952$, $e = 1.2607$& $\cO\br{T^{-2/21}}$ &$\cO\br{T^{-2/21}}$  \\ \hline
		$a = 0.8751$, $b = 0.5714$, $c = 0.7143$, $d = -0.1429$, $e = 1.1652$& $\cO\br{T^{-1/7}}$ &$\cO\br{T^{-1/14}}$ \\ \hline
		$a = 0.7143$, $b = 0.5714$, $c = 0.7143$, $d = -0.2857$, $e = 0.8518$& $\cO\br{T^{-2/7}}$ &$\cO\br{1}$  \\ \hline
	\end{tabular} 
\label{theoRes}
\end{table*}

\subsection{Performance Analysis}
This section provides the major theoretical findings of Algorithm. \ref{cscgda}. We begin our analysis by defining a new objective function by penalizing the constraint as
\begin{align}
\tilde{H}(\x,\alpha,\delta) = \tilde{f}\br{\EE\tilde{\g}\br{\x,\vxi}, \alpha,\delta} + R(\x),
\end{align}
where 
\begin{align}
\tilde{f}(\tilde{y}, \alpha,\delta) = f(\y) + \frac{\delta}{\alpha}p(\z),\hspace{3mm} p(\z) = \ell(\q(\z)) \nonumber\\  \tilde{\g}\br{\x,\vxi} = \bs{\g\br{\x,\vxi}, \h\br{\x,\vxi}}, \hspace{3mm} \tilde{\y} = \bs{\y,\z} \label{defP}. 
\end{align}
Since the inner function $\h$ is Lipschitz and has bounded gradients, It is simple to prove (see \cite[Appendix A, lemma 3]{thomdapu2019optimal}) that the function $p(\z)$ is smooth and have bounded gradient. Note that $\tilde{H}(\x,\alpha,\delta)$ is convex w.r.t $\x$. The following theorem establishes the convergence results of Algorithm. \ref{cscgda}.

\begin{theorem}\label{thm}
	Under all the assumptions in Sec.\ref{asm},  for the choice of constants which are selected as
	\begin{align*}
	\alpha_t = Ct^{-a}, \hspace{3mm} \beta_t = C_bt^{-b}, \hspace{3mm} \delta_t = Ct^{-c}, \\
	 \gamma_t = Ct^{-d}, \hspace{3mm} \eta_t = Ct^{-e},
	\end{align*}
	where $C\geq 0$, $C_b >2$, $a \geq c$, $0\leq a\leq 1$, $0\leq b\leq 1$, and $0\leq c\leq 1$, the following result holds.
	\begin{align*}
	\frac{2}{T}&\sum_{t=T/2}^T\E{H\br{\hat{\x}} - H\br{\xs}} \nonumber \\
	&\leq \cO\bigg(T^{a-1} + T^{d} + T^{4b-4c-d}+ T^{-b-d} \\
	&\hspace{1mm}+ T^{2a+4b-6c-d} + T^{2a-b-2c-d}  +T^{-2a} \\ 
	&\hspace{1mm} + T^{-a-c}+ T^{-e} + T^{2a-2c-e} + T^{e-2a} \bigg)
	\end{align*}
	\begin{align*}
	\frac{2}{T}&\sum_{t=T/2}^T\max_j\E{Q_j\br{\hat{\x}}} \\
	&\leq \cO \bigg(T^{(c-1)/2} + T^{(d+c-a)/2} + T^{(4b-3c-d-a)/2} \\
	&\hspace{1mm} + T^{(c-a-b-d)/2} + T^{(a+4b-5c-d)/2}  + T^{-a} \\
	&\hspace{1mm}+ T^{(a-b-c-d)/2} + T^{(-3a+c)/2}  + T^{(c-a-e)/2} \\
	&\hspace{1mm}+ T^{(a-c-e)/2} + T^{(e-3a+c)/2} +T^{(c-a)/2}),
	\end{align*}
	where $H(\x) = F(\x) + R(\x)$.
\end{theorem}
The proof is borrowed from  \cite[Theorem 3]{wang2017accelerating}. But the presence of stochastic constraints are need to be addressed separately from \cite{thomdapu2019optimal}. Specifically, first, we bound the absolute value of successive iterate difference as  $\E{\norm{\x_{t+1} - \x_t}^2} \leq \cO(\alpha_t^2 + \delta_t^2)$. Then we bound the difference of tracking variables and inner functions as $\E{\norm{\y_t -\bar{\g}(\x_t)}^2} \leq \cO\br{t^{-4c+4b} + t^{-b}}$ and $\E{\norm{\z_t -\bar{\h}(\x_t)}^2} \leq \cO\br{t^{-4c+4b} + t^{-b}}$. Next by deriving the bound for difference of algorithmic and optimal solutions, we prove the statement of Theorem. \ref{thm}. The complete proof is deferred to Appendix \ref{proof}.

By carefully choosing the constants, we obtain the rates, as shown in the Table \ref{theoRes}. Results provided in Theorem \ref{thm} are clearly improved convergence rates as $\cO\br{T^{-2/21}}$ compared to the best known results in \cite{thomdapu2019optimal} $\cO\br{T^{-1/12}}$. The improvement is due to making an additional smoothness assumption condition of inner functions. 

Bounds in Table \ref{theoRes} are expressed in terms of the number of iterations. The analysis excludes the per-iteration complexity, which is fixed. Aside from the reduced number of iterations to reach the optimal solution, the number of oracle calls may be more depending on the required number of queries per iteration. For example, the addition of stochastic constraints would result in requiring more gradient queries. The current approach cannot Improve the per-iteration complexity, which is the drawback of the Algorithm. \ref{cscgda}. 

\section{Simulations} \label{sim}
\begin{figure}
	\includegraphics[width=0.9\linewidth, height = 0.65\linewidth]
		{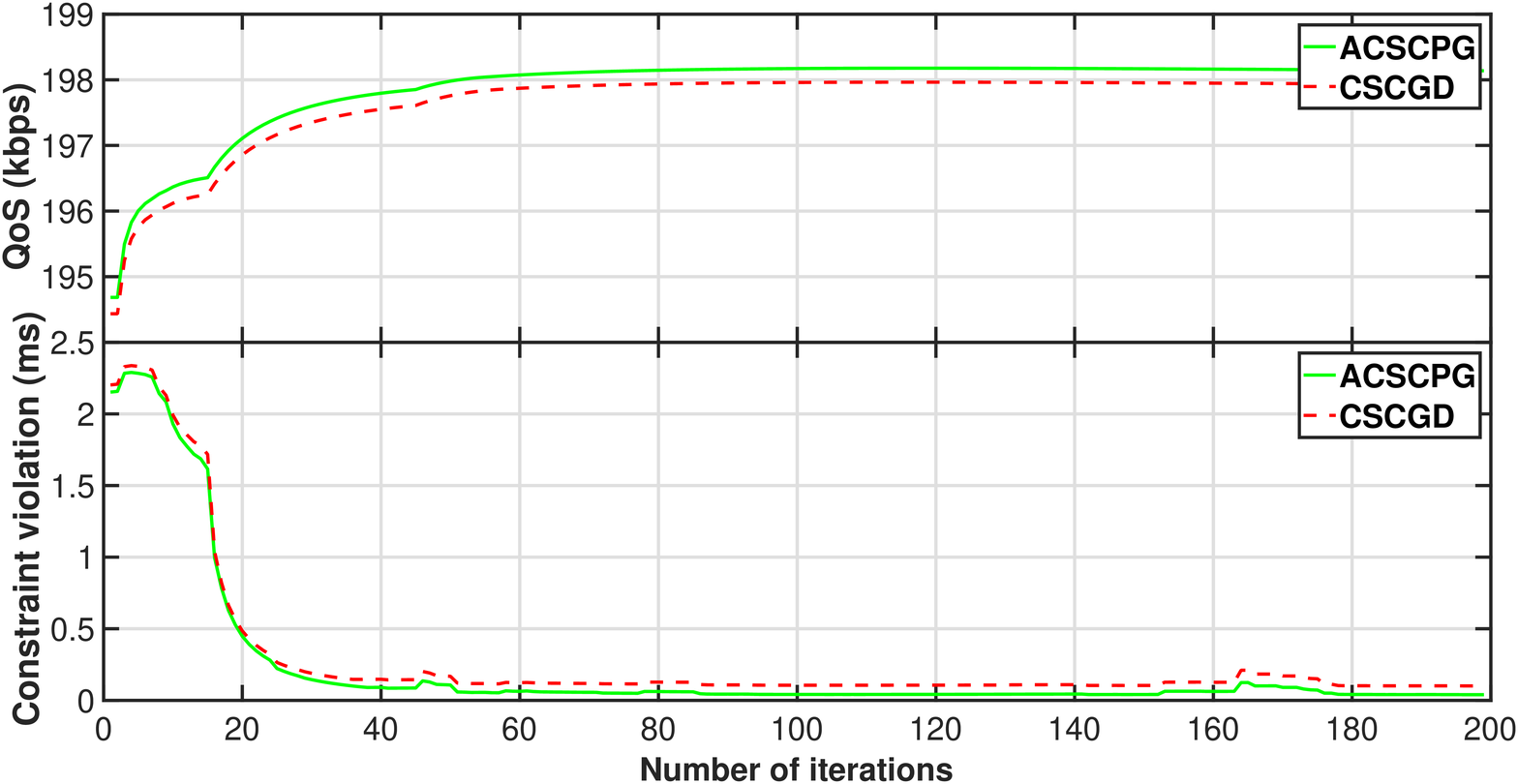}
	\caption{Convergence results for ACSCPG, CSCGD \cite{thomdapu2019optimal}}
	\label{conv-res}
\end{figure}	   
We consider a data center with $M = 10$ servers; each is equipped with a single antenna base-station capable of wireless file transmissions. The most trending files $N = 100$ are considered for placement and the popularities follow Zipf distribution. Specifically, the probability $p_k$ that $k-$th most popular file is requested at a given time adheres to $p_k \propto k^{-s}$ where $s$ is the parameter that characterizes the skewness in the distribution, which is taken as $2$ for the simulation. All the files are encoded using $(8,4)$ MDS erasure code and stored in servers, as explained in Sec. \ref{sys-mod}. Bandwidth of all the channels is allocated as $B_j = 1MBPS \text{ }\forall j=1,...,M$. Latency constraint is imposed as each server's queuing delay should not go beyond $5$ms. Experiments are conducted in the simulated environment of a unit radius circle where all the users and servers are present. The files' requests are received from various locations randomly distributed in a unit radius circle with uniform distribution. The wireless channel between users and the data center is considered Rayleigh. The attenuation factor due to the path-loss is considered to $k_0(d/d_0)^{-3}$ with the parameters $k_0=d_0=1$. Eight servers are placed on the circumference of a 0.5 unit radius circle separated with $45^{\circ}$ angular difference.  The remaining two servers are placed at the coordinates $(-2,0)$, $(2,0)$, which are far away from the users.   

As the first part of the experiments, Algorithm. \ref{cscgda} ACSCPG is run on the simulated environment to learn the policy variables $\bc{\pi_{ij}}$ and power allocations $\{p_j\}$ by trying to solve the problem \eqref{appProb} in online manner. For the comparison, CSCGD \cite{thomdapu2019optimal} is also run, and the results are shown in Fig.\ref{conv-res}. The two plots represent the evaluation of objective function $U(\p)$ (QoS metric) as in \eqref{proB}, and constraints violations $\max_j\br{W_j(p_j,\pi_ij) - D_j}$ respectively. As shown in Fig.\ref{conv-res}, the objective is maximized, while constraint violations are decreased with the number of iterations as desired. However, the proposed algorithm outperforms CSCGD in terms of convergence rate, as supported by theoretical arguments. 

\begin{table}[]
	\centering
	\begin{tabular}{|c|c|}
		\hline
		Policy & Throughput in \eqref{thr} \\ \hline
		\textbf{Equiprobable}                                                        & \textbf{178 kbps}                                                \\ \hline
		\textbf{Proposed}                                                        & \textbf{183 kbps}                                               \\ \hline
	\end{tabular}
	\caption{Comparison results for Equiprobable and the proposed policies}
	\label{toy-comp-res}
\end{table}      

To evaluate the proposed method's performance in wireless networks, we implement a heuristic technique called equiprobable policy that belongs to the category of probabilistic scheduling \cite{xiang2016joint} for the comparison.  As the name suggests, the policy variables adhere to constraints in \eqref{polfea} but have equal values. We consider average throughput that is obtained from the data center as a performance metric for the comparison and is calculated as
\begin{align} \label{thr}
	T = \frac{1}{M}\sum_{j=1}^M \frac{1}{W_j(p_j,\pi_{ij}) + \frac{1}{\sum_{i}\E{b_j(p_j,\xi_{ij})}}}.
\end{align}
A good policy should obtain higher throughput. Results are presented in Table.\ref{toy-comp-res}, and the policy obtained from the proposed method outperforms the equiprobable policy by achieving higher throughput.

\section{Conclusion and Future Scope}\label{concl}
This paper considers the erasure-coded data center in a wireless network. We propose a new scheduling policy that optimizes QoS while respecting strict queuing delay constraints. To solve the problem, we propose a new algorithm, ACSCPG, which is inspired by \cite{thomdapu2019optimal,wang2017accelerating}. Finally, we show that the proposed algorithm beats CSCGD \cite{thomdapu2019optimal} both theoretically and in a simulated environment of a data center. We also show that the policy which is obtained from the proposed approach outperforms a heuristic one called equiprobable policy.  

Apart from the benefit of rate improvements, the proposed algorithm incurs more cost per iteration with the current application. In other words, the algorithm requires approximately $M+NJ$ number of gradient queries in each iteration. The number $M+NJ$ may be much large in practice. We look for distributed versions of the current algorithm to reduce the complexity per iteration in our future works. Another interesting direction is to analyze the algorithm by considering more practical scenarios in the wireless network, such as the server nodes' mobility.

\appendices
\section{Proof of Theorem \ref{thm}} \label{proof}
Before proving the main result, we first derive some preliminary results.
\begin{lemma}\label{prilG}
	The updates in \eqref{xup} yields
	\begin{align}
	\E{\norm{\x_{t+1} - \x_t}^2} \leq \cO(\alpha_t^2 + \delta_t^2). 	
	\end{align}	
	for all $t\geq 1$. 
\end{lemma}
\begin{IEEEproof}
	From Algorithm \ref{cscgda}
	\begin{align*}
	\x_{t+1} = \pr_{\alpha_tR(.)}&\Big\{\x_t - \alpha_t\dF\g(\x_t,\vxi_t)\dF f(\y_{t}) \\
		&- \delta_t\dF \h(\x_t,\vxi_t)\dF p(\z_{t})\Big\}.
	\end{align*}
	By the definition of proximal operation, we can write
	\begin{align*}
	\x_{t+1} = \arg\min_{\x} \frac{1}{2} &\Big\|\x - \x_t - \alpha_t\dF\g(\x_t,\vxi_t)\dF f(\y_{t}) \\
	&- \delta_t\dF \h(\x_t,\vxi_t)\dF p(\z_{t})\Big\|^2 + \alpha_tR(\x).
	\end{align*}
	The optimality condition suggests,
	\begin{align*}
	\x_{t+1} -\x_t = \alpha_t\dF\g(\x_t,\vxi_t)\dF f(\y_{t}) &- \delta_t\dF \h(\x_t,\vxi_t)\dF p(\z_{t}) \\
	&+ \alpha_t\s_{t+1},
	\end{align*}
	where $\s_{t+1} = \partial R(\x_{t+1})$ norm of which is bounded. Hence,
	\begin{align*}
	\norm{\x_{t+1} -\x_t} &\leq \alpha_t\norm{\dF\g(\x_t,\vxi_t)\dF f(\y_{t})} \\
	&\hspace{5mm}+ \delta_t\norm{\dF \h(\x_t,\vxi_t)\dF p(\z_{t})} + \alpha_t\norm{\s_{t+1}} \\
	&\leq  \cO(\alpha_t + \delta_t).
	\end{align*}
\end{IEEEproof}

\begin{lemma}\label{seqL}
	Given two sequences of positive scalars $\{s_t\}_{t=1}^\infty$, and $\{\phi_t\}_{t=1}^\infty$ satisfying 
	\begin{align*}
	s_{k+1} \leq \br{1-\phi_t + C_1\phi_t^2}s_k + C_2t^{-a} + C_3t^{-c},
	\end{align*} 
	where $C_1\geq 0$, $C_2\geq 0$, $C_3\geq 0$, and $a\geq 0$ $c\geq0$. The definition $\phi_t = C_4t^{-b}$, where $b \in (0,1]$, and $C_4 > 2$, then for any $t$ the sequence is bounded as
	\begin{align*}
	s_k \leq Dt^{-d},
	\end{align*} 
	Where $D$ is
	\begin{align*}
	\max_{t\leq (C_1C_4^2)^{1/b}+1} s_tt^c + \frac{C_2+C_3}{C_4-2}, \text{ } d = \min(a-b, c-b).
	\end{align*}
\end{lemma}
\begin{IEEEproof}
	This result can be proved by induction. From the definitions it is clear that bound holds for any $t\leq (C_1C_4^2)^{1/b}$. Now assume for any $t > (C_1C_4^2)^{1/b}$, $s_t \leq Dt^{-d}$. Hence we have to prove $s_{t+1} \leq D(t+1)^{-d}$.
	\begin{align*}
	s_{t+1} &\leq \br{1-\phi_t + C_1\phi_t^2}s_k + C_2t^{-a} + C_3t^{-c} \\
	&\leq Dt^{-d} - DC_4t^{-b-d} + DC_1C_4^2t^{-2b-d} + C_2t^{-a} \\
	&\hspace{5mm}+ C_3t^{-c}.
	\end{align*}
	From the convexity of the function $f(t) = t^{-d}$, we can write
	\begin{align*}
	(t+1)^{-d} - t^{-d} \geq -dt^{-d-1}.
	\end{align*}
	To prove the result we need to prove following two steps.
	\begin{align*}
	\Delta &= (t+1)^{-d} - t^{-d} + C_4t^{-b-d} - C_1C_4^2t^{-2b-d} >0, \\ 
	D&\geq \frac{C_2t^{-a} + C_3t^{-c}}{\Delta}.
	\end{align*}
	Hence we follow
	\begin{align*}
	\Delta &\geq -dt^{-d-1} + C_4t^{-b-d} - C_1C_4^2t^{-2b-d} \\
	& \geq (C_4-2)t^{-b-d} > 0.
	\end{align*}
	The second inequality follows from the fact that $t>(C_1C_4^2)^{1/b}$, and $b\leq 1$. Finally, we consider
	\begin{align*}
	\frac{C_2t^{-a} + C_3t^{-c}}{\Delta} &\leq \frac{C_2}{C_4-2}t^{-a+b+d} + \frac{C_3}{C_4-2}t^{-c+b+d}\\
	& \leq \frac{C_2+C_3}{C_4-2}\leq D.
	\end{align*}
	The second inequality follows from the condition on $d$ which is $d = \min(a-b, c-b)$. 
\end{IEEEproof}

\begin{lemma}\label{trckL}
	If we choose $\beta_t = C_bt^{-b}$, where $C_b >2$, $b \in (0,1]$, and $\alpha_t = C_at^{-a}$, $\delta_t = C_ct^{-c}$, where $C_a\geq 0 $, $C_c\geq 0$, and $a \geq c$. Under the assumptions, we have
	\begin{align*}
	\E{\norm{\y_t -\bar{\g}(\x_t)}^2} &\leq \cO\br{t^{-4c+4b} + t^{-b}}, \\
	 \E{\norm{\z_t -\bar{\h}(\x_t)}^2} &\leq \cO\br{t^{-4c+4b} + t^{-b}}.
	\end{align*}
\end{lemma}

\begin{IEEEproof}
	We define, 
	\begin{align*}
	m_{t+1} &= \sum_{k=0}^t\zeta_k^{(t)} \norm{\x_{t+1} - \w_{k+1}}^2\\
	n_{t+1} &=  \norm{\sum_{k=0}^t\zeta_k^{(t)}\br{\g_{k+1}(\w_{k+1}) - \bar{\g}(\w_{k+1}) }}
	\end{align*}
	From  \cite[Lemma 7]{wang2017accelerating}, we have 
	\begin{align*}
	\norm{\y_t -\bar{\g}(\x_t)}^2 \leq L_g^2m_t^2 + 2n_t^2,
	\end{align*} 
	and 
	\begin{align*}
	q_{t+1} &\leq (1-\beta_t)q_k + \frac{4}{\beta_t}\norm{\x_{t+1}-\x_t}^2\\
	&\leq (1-\beta_t)q_k + \cO\br{t^{-2a+b} + t^{-2c+b}} 
	\end{align*}
	From Lemma. \ref{seqL}, we can conclude $q_k \leq \cO(t^{-2c+2b})$. Similarly from the  \cite[Lemme 7]{wang2017accelerating}, we have
	\begin{align*}
	m_{t+1} &\leq (1-\beta_t)m_t + \beta_tq_k + \frac{2}{\beta_t}\norm{\x_{t+1}-\x_t}^2 \\
	&\leq (1-\beta_t)m_t + \cO\br{t^{-2c+b}+ t^{-2a+b} + t^{-2c+b}} \\
	&\leq  (1-\beta_t)m_t +  \cO\br{t^{-2c+b}+ t^{-2a+b}}. 
	\end{align*}
	Hence we have $m_t \leq \cO(t^{-2c+2b})$. Again from the Lemma 7  in \cite{wang2017accelerating} we have
	\begin{align*}
	\E{n_{t+1}^2} \leq \br{1-2\beta_t + \beta_t^2}\E{n_{t}^2} +\cO(1)\beta_t^2.
	\end{align*}
	From the Lemma. \ref{seqL}, we have $\E{n_{t}^2} \leq \cO(t^{-b})$. Hence we conclude
	\begin{align*}
	\E{\norm{\y_t -\bar{\g}(\x_t)}^2} \leq \cO\br{t^{-4c+4b} + t^{-b}}.
	\end{align*}
	Similarly, we can also prove the other result.
\end{IEEEproof}

We have proved all the preliminary results. Now we prove the crucial result before discussing the convergence analysis.

\begin{lemma}\label{cruc}
	Under the assumptions, for any scalers $\eta_t$, and $\gamma_t$, the algorithmic updates yield
	\begin{align*}
	2&\alpha_t\E{F\br{\x_{t+1}} - F\br{\xs}} + 2\delta_t\E{P\br{\x_{t+1}} - P\br{\xs}}  \\&+ 2\alpha_t\E{R(\x_{t+1}) - R(\xs)}+ \E{\norm{\x_{t+1} - \xs}^2} \\
	&\hspace{3mm}\leq  \br{1+\frac{\alpha_t}{\gamma_t}}\E{\norm{\x_t - \xs}^2}  + \cO\br{\alpha_t^3 + \alpha_t^2\delta_t}  \\
	&\hspace{5mm} +\cO\br{L_f\alpha_t\gamma_t}\E{\norm{\y_t - \bar{\g}(\x_t)}^2}  + \cO\br{\alpha_t\eta_t + \frac{\eta_t\delta_t^2}{\alpha_t}} \\
	&\hspace{5mm}+ \cO\br{\frac{\alpha_t^3}{\eta_t}} +\cO\br{L_p\frac{\gamma_t\delta_t^2}{\alpha_t}}\E{\norm{\z_t - \bar{\h}(\x_t)}^2} 
	\end{align*} 
\end{lemma}

\begin{IEEEproof}
	Consider $\norm{\x_{t+1} - \xs}^2$
	\begin{align*}
	&= \norm{\x_{t+1} -\x_t + \x_t- \xs}^2\\
	& = \norm{\x_{t} - \xs}^2 - \norm{\x_{t+1} - \x_{t}}^2\\
	&\hspace{3mm}  + 2 \ip{\x_{t+1} - \x_t}{\x_{t+1} - \xs} \\
	&=\norm{\x_{t} - \xs}^2 - \norm{\x_{t+1} - \x_{t}}^2 \\
	&\hspace{3mm}-2\alpha_t\ip{\dF \tilde{\g}_t(\x_t)\dF \tilde{f}\br{\tilde{\y}_t,\alpha_t,\delta_t} + \s_{t+1}}{\x_{t+1} - \xs}\\
	&=\norm{\x_{t} - \xs}^2 - \norm{\x_{t+1} - \x_{t}}^2 + 2\alpha_t\ip{\s_{t+1}}{\xs-\x_{t+1}}\\
	&\hspace{3mm}+2\alpha_t\ip{\dF \tilde{\g}_t(\x_t)\dF \tilde{f}\br{\tilde{\y}_t,\alpha_t,\delta_t} }{\xs-\x_{t+1}} \\
	&\leq\norm{\x_{t} - \xs}^2 + 2\alpha_t\br{R(\xs) - R(\x_{t+1})}\\
	&\hspace{3mm} +2\alpha_t\ip{\dF \tilde{\g}_t(\x_t)\dF \tilde{f}\br{\tilde{\y}_t,\alpha_t,\delta_t} }{\xs-\x_{t+1}} \\
	&=  \norm{\x_{t} - \xs}^2  \\
	&\hspace{3mm}+ 2\alpha_t\br{T_1 +T_2} + 2\alpha_t\br{R(\xs) - R(\x_{t+1})},
	\end{align*}
	where 
	\begin{align*}
	&T_1 = \ip{\dF \tilde{F}\br{\x_t,\alpha_t,\delta_t}}{\xs-\x_{t+1}}\\
	&T_2 \\
	& = \ip{\dF \tilde{\g}_t(\x_t)\dF \tilde{f}\br{\tilde{\y}_t,\alpha_t,\delta_t} - \dF \tilde{F}\br{\x_t,\alpha_t,\delta_t} }{\xs-\x_{t+1}}.
	\end{align*}
	The inequality follows from the fact that $R(\x)$ is convex. Now consider
	\begin{align*}
	T_1 &= \ip{\dF \tilde{F}\br{\x_t,\alpha_t,\delta_t}}{\x_t-\x_{t+1}} \\
	&\hspace{3mm}+ \ip{\dF \tilde{F}\br{\x_t,\alpha_t,\delta_t}}{\xs-\x_{t}}\\
	& = \ip{\dF F\br{\x_t}}{\x_t-\x_{t+1}} +\frac{\delta_t}{\alpha_t}\ip{\dF P\br{\x_t}}{\x_t-\x_{t+1}} \\
	&\hspace{3mm}+ \ip{\dF F\br{\x_t}}{\xs-\x_{t}} + \frac{\delta_t}{\alpha_t}\ip{\dF P\br{\x_t}}{\xs-\x_{t}}\\
	&\leq F(\x_t) - F(\x_{t+1}) + \frac{\delta_t}{\alpha_t}\br{P(\x_t) - P(\x_{t+1})} \\
	&\hspace{3mm}+ \frac{1}{2}\br{L_F + \frac{\delta_t}{\alpha_t}L_P}\norm{\x_{t+1} - \x_t}^2 + F(\xs) - F(\x_t)  \\
	&\hspace{3mm} + \frac{\delta_t}{\alpha_t}\br{P(\xs) - P(\x_t)}\\
	&\leq \tilde{F}\br{\xs,\alpha_t,\delta_t} -  \tilde{F}\br{\x_{t+1},\alpha_t,\delta_t} + \cO(\alpha_t^2 + \alpha_t\delta_t).  
	\end{align*}
	Next consider $T_2$
	\begin{align*}
	 & = \ip{\dF \tilde{\g}_t(\x_t)\dF \tilde{f}\br{\tilde{\y}_t,\alpha_t,\delta_t} - \dF \tilde{F}\br{\x_t,\alpha_t,\delta_t} }{\xs-\x_{t+1}}\\
	& = T_{21} + T_{22} + \frac{\eta_t}{2}T_{23} + \frac{1}{2\eta_t}\norm{\x_t - \x_{t+1}}^2,
	\end{align*}
	where
	\begin{align*}
	T_{21} & = \Big\langle\dF \tilde{F}\br{\x_t,\alpha_t,\delta_t} - \dF\tilde{\g}_t(\x_t)\dF \tilde{f}\br{\bar{\tilde{{\g}}}(\x_t),\alpha_t,\delta_t}, \\
	&\hspace{6mm} \x_t-\xs \Big \rangle\\
	T_{22} & = \Big\langle\dF\tilde{\g}_t(\x_t)\dF \tilde{f}\br{\bar{\tilde{{\g}}}(\x_t),\alpha_t,\delta_t}  \\
	&\hspace{6mm}-\dF \tilde{\g}_t(\x_t)\dF \tilde{f}\br{\tilde{\y}_t,\alpha_t,\delta_t}, \x_t-\xs \Big\rangle \\
	T_{23} & = \norm{\dF \tilde{F}\br{\x_t,\alpha_t,\delta_t} - \dF\tilde{\g}_t(\x_t)\dF \tilde{f}\br{\tilde{\y}_t,\alpha_t,\delta_t}}^2
	\end{align*}
	We can eliminate $T_{21}$, since $\EE T_{21} = 0$. Hence we consider
	\begin{align*}
	T_{22} & \leq \frac{\gamma_t}{2}\Big\|\dF\tilde{\g}_t(\x_t)\dF \tilde{f}\br{\bar{\tilde{{\g}}}(\x_t),\alpha_t,\delta_t} \\
	&\hspace{3mm} - \dF \tilde{\g}_t(\x_t)\dF \tilde{f}\br{\tilde{\y}_t,\alpha_t,\delta_t} \Big\|^2 + \frac{1}{2\gamma_t}\norm{\x_t-\xs}^2\\
	&\leq \gamma_t\norm{\dF\g_t(\x_t)\dF f\br{\bar{\g}(\x_t)} - \dF \g_t(\x_t)\dF f\br{\y_t}}^2 \\
	&\hspace{3mm}+ \frac{\gamma_t\delta_t^2}{\alpha_t^2} \norm{\dF\h_t(\x_t)\dF p\br{\bar{\h}(\x_t)} - \dF \h_t(\x_t)\dF p\br{\z_t}}^2\\
	&\hspace{3mm} + \frac{1}{2\gamma_t}\norm{\x_t-\xs}^2\\
	&\leq \cO\br{L_f\gamma_t}\norm{\y_t - \bar{\g}(\x_t)}^2 + \frac{1}{2\gamma_t}\norm{\x_t-\xs}^2 \\
	&\hspace{3mm} + \cO\br{L_p\frac{\gamma_t\delta_t^2}{\alpha_t^2}}\norm{\z_t - \bar{\h}(\x_t)}^2.		
	\end{align*}
	Finally we consider
	\begin{align*}
	T_{23} & = \norm{\dF \tilde{F}\br{\x_t,\alpha_t,\delta_t} - \dF\tilde{\g}_t(\x_t)\dF \tilde{f}\br{\tilde{\y}_t,\alpha_t,\delta_t}}^2\\
	&= \Big\|\dF F\br{\x_t} + \frac{\delta_t}{\alpha_t}\dF P(\x_t) -  \dF \g_t(\x_t)\dF f\br{\y_t} \\
	&\hspace{3mm}- \frac{\delta_t}{\alpha_t} \dF \h_t(\x_t)\dF p\br{\z_t}\Big\|^2 \\
	&\leq \cO\br{1 + \frac{\delta_t^2}{\alpha_t^2}}.
	\end{align*}
	Now By considering all the intermediate results we conclude
	\begin{align*}
	&\E{\norm{\x_{t+1} - \xs}^2} \leq \br{1+\frac{\alpha_t}{\gamma_t}}\E{\norm{\x_t - \xs}^2} \\
	& + 2\alpha_t\tilde{F}\br{\xs,\alpha_t,\delta_t} -  2\alpha_t\E{\tilde{F}\br{\x_{t+1},\alpha_t,\delta_t}} + \cO\br{\frac{\alpha_t^3}{\eta_t}}  \\
	&+\cO\br{L_f\alpha_t\gamma_t}\E{\norm{\y_t - \bar{\g}(\x_t)}^2}  + \cO\br{\alpha_t\eta_t + \frac{\eta_t\delta_t^2}{\alpha_t}}  \\
	& + \cO\br{L_p\frac{\gamma_t\delta_t^2}{\alpha_t}}\E{\norm{\z_t - \bar{\h}(\x_t)}^2} + \cO(\alpha_t^3 + \alpha_t^2\delta_t) \\
	&+ 2\alpha_t\E{R(\xs) - R(\x_{t+1})} .
	\end{align*}
	We get the required result by interchanging the terms.
\end{IEEEproof}

Now we are ready with all the derived results, we proceed to prove the theorem. Let us denote 
\begin{align*}
H(\x) = F(\x) + R(\x).
\end{align*}
Further we know $P(\x^{\star}) = 0$. Now by summing over $1,...,T$ the expression of Lemma. \ref{cruc}, we get
\begin{align*}
&\sum_{t=T/2}^T\br{2\E{H\br{\x_{t+1}} - H\br{\xs}} + 2\frac{\delta_t}{\alpha_t}\E{P\br{\x_{t+1}}} } \\
&\hspace{3mm}\leq \frac{1}{\alpha_{T/2}}\E{\norm{\x_1 - \xs}^2} + \sum_{t=T/2}^T\frac{1}{\gamma_t}\E{\norm{\x_t - \xs}^2} \\
&\hspace{5mm}+\sum_{t=T/2}^T\cO\br{\frac{\gamma_t\delta_t^2}{\alpha_t^2}}\E{\norm{\z_t - \bar{\h}(\x_t)}^2}  \\
&\hspace{5mm}+\sum_{t=T/2}^T\cO\br{\gamma_t}\E{\norm{\y_t - \bar{\g}(\x_t)}^2}\\
& \hspace{5mm} + \sum_{t=T/2}^T\cO\br{\alpha_t^2 + \alpha_t\delta_t + \eta_t + \frac{\eta_t\delta_t^2}{\alpha_t^2} + \frac{\alpha_t^2}{\eta_t}} \\
&\hspace{3mm}\leq  \sum_{t=T/2}^T\cO\br{\frac{\gamma_t\delta_t^2}{\alpha_t^2}}\E{\norm{\z_t - \bar{\h}(\x_t)}^2}  +\cO\br{\frac{1}{\alpha_{T/2}}} \\
& \hspace{5mm} + \cO\br{\sum_{t=T/2}^T\frac{1}{\gamma_t}} + \sum_{t=T/2}^T\cO\br{\gamma_t}\E{\norm{\y_t - \bar{\g}(\x_t)}^2} \\
& \hspace{5mm} + \sum_{t=T/2}^T\cO\br{\alpha_t^2 + \alpha_t\delta_t + \eta_t + \frac{\eta_t\delta_t^2}{\alpha_t^2} + \frac{\alpha_t^2}{\eta_t}}
\end{align*}
The second inequality follows from the fact that $\norm{\x_t-\xs}^2 \leq \cO(1)$. From the results in Lemma. \ref{trckL}, we can write
\begin{align}
&\sum_{t=T/2}^T\br{2\E{H\br{\x_{t+1}} - H\br{\xs}} + 2\frac{\delta_t}{\alpha_t}\E{P\br{\x_{t+1}}} } \nonumber\\
&\hspace{3mm}\leq  \sum_{t=T/2}^T\br{\cO\br{\gamma_t} +   \cO\br{\frac{\gamma_t\delta_t^2}{\alpha_t^2}}  } \cO\br{\br{\frac{\delta_t}{\beta_t}}^4 + \beta_t}  \nonumber\\
&\hspace{5mm}+ \sum_{t=1}^T\cO\br{\alpha_t^2 + \alpha_t\delta_t + \eta_t + \frac{\eta_t\delta_t^2}{\alpha_t^2} + \frac{\alpha_t^2}{\eta_t}} \nonumber\\
&\hspace{5mm}+\cO\br{\frac{1}{\alpha_{T/2}}} + \cO\br{\sum_{t=T/2}^T\frac{1}{\gamma_t}}\nonumber\\
&\hspace{3mm}\leq \sum_{t=T/2}^T\cO\br{ \frac{\gamma_t\delta_t^4}{\beta_t^4} + \beta_t\gamma_t + \frac{\gamma_t\delta_t^6}{\alpha_t^2\beta_t^4} + \frac{\gamma_t\delta_t^2\beta_t}{\alpha_t^2}} \nonumber\\
&\hspace{5mm} + \sum_{t=T/2}^T\cO\br{\alpha_t^2 + \alpha_t\delta_t + \eta_t + \frac{\eta_t\delta_t^2}{\alpha_t^2} + \frac{\alpha_t^2}{\eta_t}} \nonumber\\
&\hspace{5mm}+\cO\br{\frac{1}{\alpha_{T/2}}} + \cO\br{\sum_{t=T/2}^T\frac{1}{\gamma_t}} \label{intrV}
\end{align}
Now For the first result since we know $P\br{\x_{t+1}} \geq 0$, we write
\begin{align*}
&\sum_{t=T/2}^T2\E{H\br{\x_{t+1}} - H\br{\xs}} \\
&\hspace{3mm}\leq \sum_{t=T/2}^T\cO\br{ \frac{\gamma_t\delta_t^4}{\beta_t^4} + \beta_t\gamma_t + \frac{\gamma_t\delta_t^6}{\alpha_t^2\beta_t^4} + \frac{\gamma_t\delta_t^2\beta_t}{\alpha_t^2}} \\
&\hspace{5mm} + \sum_{t=T/2}^T\cO\br{\alpha_t^2 + \alpha_t\delta_t + \eta_t + \frac{\eta_t\delta_t^2}{\alpha_t^2} + \frac{\alpha_t^2}{\eta_t}} \\
&\hspace{5mm}+\cO\br{\frac{1}{\alpha_{T/2}}} + \cO\br{\sum_{t=T/2}^T\frac{1}{\gamma_t}}.
\end{align*}
After substituting all the constants choices, 
\begin{align*}
&\sum_{t=T/2}^T2\E{H\br{\x_{t+1}} - H\br{\xs}} \leq \cO\br{T^{a}} + \cO\br{\sum_{t=T/2}^T t^{d}} \\
& +\sum_{t=T/2}^T\cO\Big( t^{4b-4c-d} + t^{2a+4b-6c-d} + t^{2a-b-2c-d}  \\
& + t^{-b-d} + t^{-2a} + t^{-a-c} + t^{-e} + t^{2a-2c-e} + t^{e-2a} \Big) \\
&\approxeq \cO\Big(T^{a} + T^{d+1} + T^{1+4b-4c-d}+ T^{1-b-d}   \\
&+ T^{1+2a+4b-6c-d} + T^{1+2a-b-2c-d} + T^{1-2a} + T^{1-a-c}  \\
&+ T^{1-e}+ T^{1+2a-2c-e} + T^{1+e-2a} \Big) .
\end{align*}
Proceeding further
\begin{align*}
&\frac{2}{T}\sum_{t=T/2}^T\E{H\br{\x_{t+1}} - H\br{\xs}} \leq \cO\Big( T^{a-1} + T^{d} \\
&\hspace{3mm}+ T^{4b-4c-d}+ T^{-b-d} + T^{2a+4b-6c-d} + T^{2a-b-2c-d}  \\
&\hspace{3mm} +\cO T^{-2a} + T^{-a-c} + T^{-e} + T^{2a-2c-e} + T^{e-2a}  \Big ).
\end{align*}
Hence from the convexity of $H$, first result of Thm.\ref{thm} is proved.

Now for the second results since we know $|\E{H\br{\x_{t+1}} - H\br{\xs}} | \geq -\cO(1)$. Since $\alpha_t \leq \delta_t$, and the function $P$ is convex we have
\begin{align}
\frac{T\delta_{T/2}}{2\alpha_{T/2}}\E{P\br{\hat{\x}}} 	&\leq  \frac{\delta_{T/2}}{\alpha_{T/2}}\sum_{t=T/2}^T\E{P\br{\x_{t+1}}} \nonumber \\
&\leq \sum_{t=T/2}^T\frac{\delta_t}{\alpha_t}\E{P\br{\x_{t+1}}} \label{lowe}. 
\end{align}
From the definition of function $P$ provided in \eqref{defP}, and penalty function in $\eqref{hub}$, we know $P(\hat{\x}) = \sum_j\br{\bs{Q_j(\hat{\x})}_+}^2$ where $[.]_+$ is the projection on positive orthant. Therefore the expression in \eqref{lowe} becomes
\begin{align} 
\frac{T\delta_{T/2}}{2J\alpha_{T/2}}\br{\sum_{j=1}^J\bs{Q_j(\hat{\x})}_+}^2 &\leq \frac{T\delta_{T/2}}{2\alpha_{T/2}}\sum_{j=1}^J\br{\bs{Q_j(\hat{\x})}_+}^2 \nonumber \\
&\leq \sum_{t=T/2}^T\frac{\delta_t}{\alpha_t}\E{P\br{\hat{\x}}}\label{conFi}.
\end{align} 
From the expression in \eqref{intrV}, we can write
\begin{align*}
&2\frac{\delta_{T/2}}{\alpha_{T/2}}\sum_{t=T/2}^T\E{P\br{\x_{t+1}}} \\
&\hspace{3mm}\leq \sum_{t=T/2}^T\cO\br{ \frac{\gamma_t\delta_t^4}{\beta_t^4} + \beta_t\gamma_t + \frac{\gamma_t\delta_t^6}{\alpha_t^2\beta_t^4} + \frac{\gamma_t\delta_t^2\beta_t}{\alpha_t^2}}\\
&\hspace{5mm} + \sum_{t=T/2}^T\cO\br{\alpha_t^2 + \alpha_t\delta_t + \eta_t + \frac{\eta_t\delta_t^2}{\alpha_t^2} + \frac{\alpha_t^2}{\eta_t}} \\
&\hspace{5mm}+\cO\br{\frac{1}{\alpha_{T/2}}} + \cO\br{\sum_{t=T/2}^T\frac{1}{\gamma_t}} + \cO(T) \\
&\hspace{3mm}\approxeq \cO \Big(T^{a} + T^{d+1} + T^{1+4b-4c-d} + T^{1+2a+4b-6c-d} \\
&\hspace{5mm} + T^{1-b-d} + T^{1+2a-b-2c-d} + T^{1-2a} + T^{1-a-c}  \\
&\hspace{5mm} + T^{1-e} + T^{1+2a-2c-e} + T^{1+e-2a} +T \Big). 
\end{align*}
By taking $\delta_{T/2}/\alpha_{T/2}$ to R.H.S and dividing with $T$, we can write
\begin{align*}
&\frac{2}{T}\sum_{t=T/2}^T\E{P\br{\x_{t+1}}} \nonumber \\
&\hspace{3mm}\leq \cO\Big(T^{c-1} + T^{d+c-a} + T^{4b-3c-d-a}+ T^{c-a-b-d}   \\
&\hspace{5mm}+ T^{a+4b-5c-d} + T^{a-b-c-d}+T^{-3a+c} + T^{-2a} \\
&\hspace{5mm} + T^{c-a-e} + T^{a-c-e} + T^{e-3a+c} +T^{c-a} \Big)
\end{align*}
From the expression in \eqref{conFi}, we conclude the proof by saying

\begin{align*}
&\sum_{j=1}^J\bs{Q_j(\hat{\x})}_+ \\
&\hspace{3mm}\leq \cO\Big(T^{(c-1)/2} + T^{(d+c-a)/2} + T^{(4b-3c-d-a)/2} \\
&\hspace{5mm} + T^{(c-a-b-d)/2} + T^{(a+4b-5c-d)/2} + T^{(a-b-c-d)/2} \\
&\hspace{5mm} +T^{(-3a+c)/2} + T^{-a} + T^{(c-a-e)/2} + T^{(a-c-e)/2} \\
&\hspace{5mm} + T^{(e-3a+c)/2} +T^{(c-a)/2}\Big).
\end{align*}

 \footnotesize
    \bibliographystyle{IEEEtran}
    \bibliography{IEEEabrv,references}

\end{document}